\documentclass[prb, amsmath,twocolumn, amssymb,superscriptaddress, floatfix]{revtex4-2}

\usepackage{amsmath}
\usepackage{amssymb}
\usepackage{amsthm}
\usepackage{mathtools}
\usepackage[pdftex]{color}
\usepackage{graphicx}% Include figure files
\usepackage{dcolumn} % Align table columns on decimal point
\usepackage{bm} % bold math
\usepackage{epic}
\usepackage{longtable}
\usepackage{ulem}   % to strike things out
\newcommand{\HH}{\mathcal{H}}
\newcommand{\kk}{\textbf{k}}

\newcommand{\q}{\textbf{q}}

\newcommand{\tr}{{\rm tr}}
\newcommand{\pdag}{{\phantom{\dag}}}

\newcommand{\sqrta}{\sqrt{|\vec{f}_{\kk}|^2 + |\vec{\xi}_{\kk}^-|^2}}

\begin{document}
\title{
    %Superconductivity in a staggered Rashba system: from two to three dimensions     
  Spin response and topology of a staggered Rashba superconductor
}
\author{
    Anastasiia~Skurativska
}
\affiliation{
   Department of Physics, University of Zurich, Winterthurerstrasse 190, 8057 Zurich, Switzerland
}
\author{
    Manfred Sigrist
}
\affiliation{
    Institute for Theoretical Physics, ETH Zurich, 8093 Zurich, Switzerland
}
\author{
    Mark H.~Fischer
}
\affiliation{
   Department of Physics, University of Zurich, Winterthurerstrasse 190, 8057 Zurich, Switzerland
}
\date{\today}

\begin{abstract}
    Inversion symmetry is a key symmetry in unconventional superconductors and even its local breaking can have profound implications.
    For inversion-symmetric systems, there is a competition on a microscopic level between  the spin-orbit coupling associated with the local lack of inversion and hybridizing terms that `restore' inversion.
    Investigating a layered system with alternating mirror-symmetry breaking, we study this competition considering the spin response of different superconducting order parameters for the case of strong spin-orbit coupling.
    We find that signatures of the local non-centrosymmetry, such as an increased spin susceptibility in spin-singlet superconductors for $T\rightarrow 0$,  persist even into the quasi-three-dimensional regime. This leads to a direction dependent spin response which allows to distinguish different superconducting order parameters.
    Furthermore, we identify several regimes with possible topological superconducting phases within a symmetry-indicator analysis. 
    Our results may have direct relevance for the recently reported Ce-based superconductor CeRh$_2$As$_2$ and beyond.
\end{abstract}

\maketitle

\section{Introduction}
In superconductors with inversion symmetry, even- and odd-parity gap functions are distinct by symmetry and, thus, correspond by the Pauli principle also to spin-singlet and spin-triplet pairing states, respectively. As a consequence, magnetic response can be used to distinguish the two cases.
When the system lacks inversion, however, spin-singlet and spin-triplet can mix~\cite{smidman:2017}. In addition, the spin-orbit coupling associated with the broken symmetry---an example is the spin-orbit coupling of Rashba type for broken in-plane mirror symmetry---strongly restricts the possible spin-triplet order-parameter components, in other words it fixes the direction of the $d$ vector~\cite{anderson:1984}. Moreover, even in the case of dominant spin-singlet or spin-triplet order parameters, the magnetic response, such as the spin susceptibility or critical fields, is not a feasible distinguishing probe anymore, as for spin-singlet superconductors, a finite spin susceptibility for $T\rightarrow 0$ and unusually high critical fields can be expected. Beyond that, it was shown that topological properties of the phases, such as vortex bound states or surface flat bands, can serve as fingerprints of the respective phases~\cite{lu:2008, schnyder:2011}.

Signatures of non-centrosymmetry can survive even in inversion-symmetric systems~\cite{fischer:2011b, sigrist:2014}. In particular, a crystal comprising weakly-coupled sublattices whose subunits locally lack inversion, such as in the hexagonal SrPtAs~\cite{youn:2012} or even in some high-temperature cuprates~\cite{gotlieb:2018}, can exhibit an unconventional magnetic response or intriguing spin textures. In addition to specific crystal structures, this {\it local non-centrosymmetricity} can arise in artificial superlattices such as the regular stacks of superconducting CeCoIn$_5$ alternating with layers of YbCoIn$_5$~\cite{shishido:2010, mizukami:2011, maruyama:2012b}. Note that the effect of local non-centrosymmetricity depends on the relative strength of inversion-breaking-induced spin-orbit coupling and inter-sublattice hybridization. The focus of most studies has, thus, been on quasi-two-dimensional systems with weak $c$-axis dispersion.

The recently discovered heavy-Fermion superconductor CeRh$_2$As$_2$ with its tetragonal crystal structure belongs also to the class of locally non-centrosymmetric superconductors. In particular, it consists of layers with alternating inversion-symmetry breaking. The upper critical field directed along the $c$ axis (perpendicular to the staggered layers) extrapolates to  $\sim 14$ T at zero temperatures, which lies far beyond the paramagnetic limiting field $H_p \sim 0.5$ T for a critical temperature $T_c \approx 0.26$K~\cite{khim:2021tmp}. Furthermore, the upper critical field shows a pronounced kink for a field $H\approx 4$ T. This anomaly strongly suggests a change in the order-parameter symmetry upon increasing magnetic field as also found in recent Ginzburg-Landau studies~\cite{schertenleib:2021tmp, moeckli:2021tmp}. Note that the critical field for in-plane directions, on the other hand, extrapolates to only approximately $2$T.

Unlike most staggered systems studied so far, CeRh$_2$As$_2$ is expected to have a rather strong $c$-axis dispersion, in other words it is a three-dimensional (3D) system. Being a Ce-based superconductor, we expect also a sizable spin-orbit coupling. Motivated by these observations, we revisit the physics of locally non-centrosymmetric superconductors in situations where both, the inter-sublattice hopping and the spin-orbit coupling strength are comparable to each other and the overall band width.

In particular, we investigate in detail a microscopic model of a layered system, where mirror symmetry is broken in a staggered fashion. By design, our model allows us to investigate how the system evolves from the 2D limit to the truly 3D case. For this purpose, we first study the spin susceptibility in the normal state, which leads us to identify four distinct regions, going from quasi-two-dimensional (q2D) all the way to truly 3D. Then we discuss possible order parameters and analyze their spin response, showing how both in-plane and out-of-plane fields are necessary to distinguish them. Eventually, we address the topological phases in the fully gapped case. Note that since the system retains inversion, we can use the recently developed concept of symmetry indicators~\cite{skura:2020, ono:2020, geier:2020, huang:2020tmp} to show that the system can realize both first- and second-order topological superconducting phases.

\section{Normal State Properties}
\label{sec:normal}
\subsection{Microscopic model}
\begin{figure}[bt]
  \centering
  \includegraphics{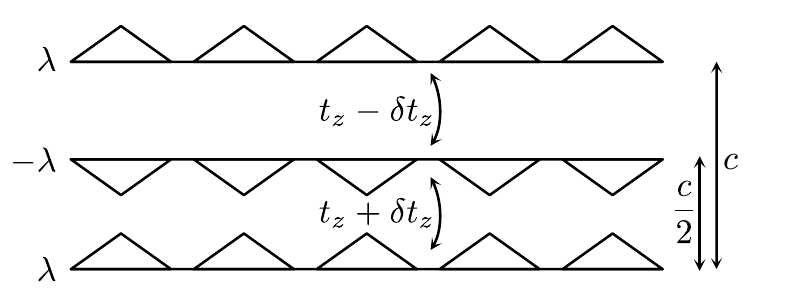}
  \caption{Schematic of the system with the triangles indicating the mirror symmetry breaking. Note that, for simplicity, we choose the layers to be spaced equidistantly in the $z$ direction with unit-cell size $c$.}
  \label{fig:setup}
\end{figure}

We consider in the following a system of stacked layers, where each layer lacks a mirror symmetry in such a way that the resulting Rashba spin-orbit coupling alternates in sign, see Fig.~\ref{fig:setup}. 
Such a system has centers of inversion, which lie in between two neighboring layers.
However, electrons moving within an individual layer are subject to a spin-orbit coupling of Rashba type. 
For concreteness, we choose these layers to consist of a square lattice with point group symmetry $C_{4v}$, while the full structure involves the tetragonal point group $D_{4h}$.
On a single layer, the electrons are governed by the Hamilontian
\begin{equation}
  \HH_{\rm sl} = \sum_{\kk s} \xi_{\kk}^+ c_{\kk s}^\dag c_{\kk s}^{\pdag} + \sum_{\kk s s'}\vec{f}_{\kk}\cdot\vec{\sigma}_{ss'} c_{\kk s}^{\dag} c_{\kk s'}^{\pdag},
  \label{eq:single_layer}
\end{equation}
where $c^\dag_{\kk s}$ ($c^\pdag_{\kk s}$) creates (annihilates) an electron with momentum $\kk$ and spin $s$. 
Setting the in-plane lattice constant $a=1$, 
\begin{equation}
  \xi_{\kk}^+ = - 2 t (\cos k_x + \cos k_y)
  \label{eq:nn_hopping}
\end{equation}
describes the dispersion due to nearest-neighbor hopping,
\begin{align}
  f_{\kk}^x &=  \lambda \sin k_y\\
  f_{\kk}^y &= -\lambda \sin k_x
\end{align}
enters the expression for the Rashba spin-orbit coupling, and $\vec{\sigma}$ denote the Pauli matrices. 

The three-dimensional system with a staggered stacking of such layers, as shown in Fig.~\ref{fig:setup}, is then described by the Hamiltonian
\begin{equation}
  \HH = \sum_{\kk} \psi_{\kk s}^{\dag}\HH_{\kk} \psi^{\phantom{\dag}}_{\kk s},
  \label{eq:ham0}
\end{equation}
with the $4\times 4$ matrix
\begin{equation}
  \HH_{\kk} = \xi_{\kk}^+\sigma_0\tau_0 + \vec{\xi}_{\kk}^-\cdot\sigma_0\vec{\tau} + \vec{f}_{\kk}\cdot\vec{\sigma}\tau_{3},
  \label{eq:hamks}
\end{equation}
where the inter-layer hopping is given by
\begin{align}
  (\xi_{\kk}^-)_1 &= - 2 t_z \cos (k_z/2)\label{eq:inter_layer1}\\
  (\xi_{\kk}^-)_2 &= - 2 \delta t_z \sin (k_z/2)
  \label{eq:inter_layer2}
\end{align}
and $(\xi_{\kk}^-)_3 = 0$. Note that we have set the $z$-axis lattice constant $c=1$.
Furthermore, we have introduced the Pauli matrices $\tau_i$, $i=1,2,3$, acting on the sublattice space of layers for the operators $\psi_{\kk s}^\dag = (c^{\dag}_{{\rm e}\kk s}, c^{\dag}_{{\rm o}\kk s})$ denoted by even (e) and odd (o) sublattice index.

The eigenenergies of this Hamiltonian are doubly degenerate and are given by
\begin{equation}
  \xi_{\kk\pm} = \xi_{\kk}^+ \pm\sqrt{|\vec{\xi}_{\kk}^{-}|^2 + |\vec{f}_{\kk}|^2} - \mu, %\equiv \xi_{\alpha},
  \label{eq:energies}
\end{equation}
which we denote in the following as $ \xi_{\alpha} $ with $\alpha = \pm$ neglecting the momentum index for shorter notation. Further, we introduce the chemical potential $\mu$.
For concreteness, we use in the following $\lambda = 0.5 t$ with $t$ the energy unit, and we choose $\mu$ so as to fix the density of electrons to $n_{\rm tot} = 0.15$.

\subsection{Magnetic Response}
It is instructive to first consider the normal-state magnetic response of this system.
For this purpose, we introduce the normal state Green's function, which is defined through
\begin{equation}
  G_0(\kk, \omega_n)^{-1} = i\omega_n\sigma_0\tau_0 - \HH_{\kk}
  \label{eq:g0inv}
\end{equation}
with $\omega_n = (2n+1)\pi T$ the fermionic Matsubara frequencies with $n\in \mathbb{Z}$ and $T$ the temperature.
This $4\times 4$ Green's--function matrix can explicitly be calculated by inverting Eq.~\eqref{eq:g0inv},
\begin{equation}
  G_0(\kk, \omega_n)=G^0_{+}\sigma_0\tau_0 + G^0_{-}(\hat{\xi}_{\kk}^{-}\cdot\sigma_0\vec{\tau} + \hat{f}_{\kk}\cdot\vec{\sigma}\tau_{3}),
  \label{eq:normalgreens0}
\end{equation}
where we introduced
\begin{equation}
  G^0_{\pm} \equiv G^0_{\pm}(\kk, \omega_n) = \frac12\Big(\frac{1}{i\omega_n - \xi_+}\pm \frac{1}{i\omega_n - \xi_-}\Big),
  \label{eq:normalgpm0}
\end{equation}
\begin{equation}
  \hat{\xi}_{\kk}^{-} = \frac{\vec{\xi}_{\kk}^{-}}{\sqrt{|\vec{f}_{\kk}|^2 + |\vec{\xi}_{\kk}^{-}|^2}},
  \label{eq:xihat}
\end{equation}
and 
\begin{equation}
  \hat{f}_{\kk} = \frac{\vec{f}_{\kk}}{\sqrt{|\vec{f}_{\kk}|^2 + |\vec{\xi}_{\kk}^{-}|^2}}.
  \label{eq:fhat}
\end{equation}

In order to investigate the system's response to a magnetic field, we calculate the (normal-state) uniform, static spin susceptibility ($\q \!=\! 0, \, \omega \!=\! 0$), which reads
\begin{equation}
  \chi^0_{ij} = -\mu_B^2 T \sum_n\sum_{\kk}\tr[\sigma_i\tau_0G_0(\kk, \omega_n)\sigma_{j}\tau_0G_0(\kk, \omega_n)],
  \label{eq:chi0}
\end{equation}
where the trace runs over spin and layer indices.
Performing the trace first, we find that due to spin-orbit coupling the susceptibility has a generic form with two contributions: the first is
\begin{equation}
\begin{split}
  \chi^{0}_{\rm P}(\kk) &= -4 \mu_{B}^{2}T\sum_{\omega_n}[(G^{0}_{+})^{2} + (G^{0}_{-})^{2}]\\ 
  &= 2\mu_{B}^{2}\Big[\frac{\partial n_{\rm F}(\xi_{+})}{\partial \xi_{+}} + \frac{\partial n_{\rm F}(\xi_{-})}{\partial \xi_{-}}\Big]\propto S_{\kk}(\mu)
  \label{eq:sum1}
  \end{split}
\end{equation}
with $n_{\rm F}(\xi)$ the Fermi distribution.
This term corresponds to the Pauli-like susceptibility. In other words, it is proportional to the spectral density at the Fermi level, $S_{\kk}(\mu)$, and after the $k$-integration, the density of states of the two bands at the Fermi level, $N(\mu)$. The second is
\begin{equation}
\begin{split}
  \chi^{0}_{\rm vV}(\kk) &=  -4 \mu_{B}^{2}T\sum_{\omega_n}[(G^{0}_{+})^{2} - (G^{0}_{-})^{2}]\\
  &= 2\mu_{B}^2\frac{n_{\rm F}(\xi_{+})-n_{\rm F}(\xi_{-})}{\sqrt{|\vec{f}_{\kk}|^2 + |\vec{\xi}_{\kk}^-|^2}}
  \label{eq:sum2}
  \end{split}
\end{equation}
and originates from inter-band processes due to the spin-orbit coupling and thus, we refer to it as a van Vleck term.
The susceptibility is a combination of these Pauli and van Vleck contributions, in particular
\begin{equation}
\chi^{0}_{z} = \sum_{\kk}|\hat{\xi}_{\kk}^-|^2\chi^{0}_{\rm P}(\kk) + |\hat{f}_{\kk}|^2\chi^{0}_{\rm vV}(\kk)
  \label{eq:chiznormal}
\end{equation}
for fields in the $z$ direction, $i=j=z$. For fields along the $x$ direction, we find 
\begin{equation}
  \chi_{x}^{0} = \sum_{\kk}[|\hat{\xi}_{\kk}^-|^2 + (\hat{f}_\kk^x)^2]\chi^{0}_{\rm P}(\kk) + (\hat{f}_\kk^y)^2\chi^{0}_{\rm vV}(\kk)
  \label{eq:chixnormal}
\end{equation}
and analogously for fields in $y$ direction. This result generalizes the pure Rashba case~\cite{smidman:2017, frigeri:2004}, which is recovered when setting $t_z = \delta t_z = 0$. In the other limit, $t_z, \delta t_z \gg \lambda$, the weight of the van Vleck term is strongly reduced and the Pauli susceptibility dominates. Consequently, we can use the ratio of van Vleck susceptibility to the total susceptibility for fields along the $z$ axis as a measure of the inversion-symmetry breaking in the system.

\begin{figure}[bt]
  \centering
  \includegraphics{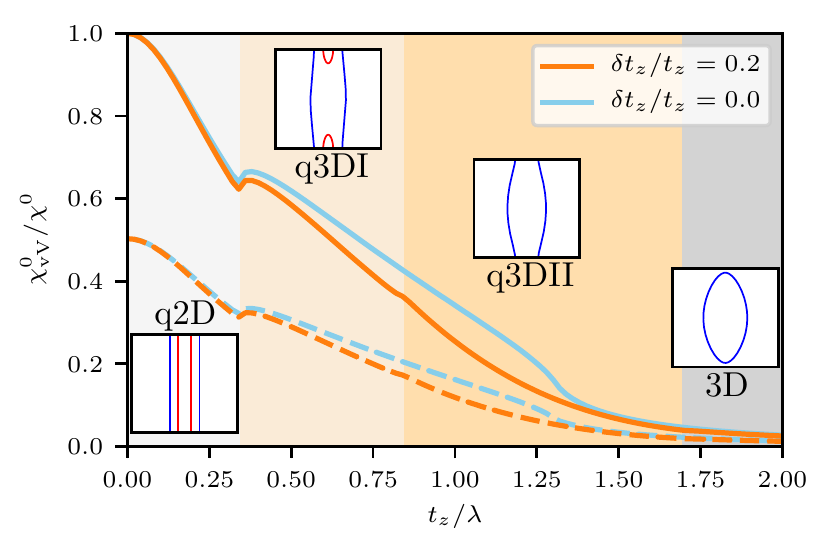}
  \caption{Ratio of van Vleck contribution to the total susceptibility in the normal state for fields along the $z$ direction (solid lines) and in plane (dashed lines). For the case of $\delta t_z/t_z = 0.2$, four different regimes are indicated, which are connected to the Fermi surface topology, see insets. For numerical reasons, we set $T=0.0025 t$.}
  \label{fig:chi0}
\end{figure}

Figure~\ref{fig:chi0} shows the ratio of van Vleck susceptibility to total susceptibility of the normal state as a function of the interlayer hopping $t_z$. The susceptibility shows a distinct behavior depending on the Fermi surface topology as indicated by the insets: In the quasi-two-dimensional (q2D) regime, where both bands are partially filled and with open Fermi surfaces, 
the van Vleck contribution drops rapidly with increasing $t_z$. Then, in the first quasi-3D regime (q3DI) with one closed
Fermi surface, $ \chi_{\rm vV}^0/\chi^0 $ decreases essentially
linearly, before dropping again more rapidly once the closed Fermi surface disappears in the second quasi-3D regime (q3DII). Finally, the remaining Fermi surface closes, at which point the full system is 3D and behaves completely centrosymmetric. As we will discuss in Sec.~\ref{sec:topology}, the different Fermi-surface topologies are also connected to different possible topological phases for a triplet superconductor.

Before we move to calculating the spin susceptibility in the superconducting state, we can use the above finding to discuss what we expect for a spin-singlet gap function. As the Pauli term depends on the density of states at the Fermi level, which vanishes for a full (spin-singlet) gap, the corresponding susceptibility is expected to vanish as well. The van Vleck term, however, should stay constant, at least for $\sqrta \gg \Delta$, with $\Delta$ the superconducting gap. Comparing Eqs.~\eqref{eq:chiznormal} and \eqref{eq:chixnormal}, we further expect the susceptibility for fields in-plane to be half the size of the susceptibility for fields along $z$ for $T\rightarrow0$, independent of the microscopic details, as long as the system has $C_4$ symmetry.

\section{Spin response of the superconductor}
\label{sec:SC}
\subsection{Superconducting order parameters}
Before calculating the spin response of possible superconducting phases, we discuss the possible pairing states of interest here. While mixing of spin-singlet and spin-triplet order parameters is allowed within the plane and microscopically supported by the Rashba spin-orbit coupling, the order parameter can still be classified as even or odd due to the global inversion symmetry~\cite{fischer:2011b}. 
Similar to the non-centrosymmetric situation~\cite{smidman:2017, anderson:1984, frigeri:2004}, most spin-triplet order parameters are suppressed by the spin-orbit coupling in the staggered case, too~\cite{fischer:2011b}.
The energetically most stable gap structures correspond to intra-band pairing, in other words to gap functions that are diagonal in the bands with energies given by Eq.~\eqref{eq:energies}~\cite{fischer:2013b, ramires:2016}. As these order parameters can also be treated analytically, we will focus exclusively on them. As a consequence, the order parameters we study have no in-plane mixing of spin-singlet and spin-triplet channels. Note that we assume in the following the superconducting gap to be the smallest energy scale in the problem.

For this intra-band pairing, we focus on the most symmetric gap functions leading to three different gap structures: a spin-singlet that transforms like  $A_{1g}$, a spin-triplet order parameter with its $d$-vector in plane that transforms like  $A_{2u}$, and a spin-triplet order parameter with the corresponding $d$-vector along the $z$ axis, that transforms like $A_{1u}$ in $D_{4h}$~\cite{fischer:2011b}. We briefly discuss in the following all three order parameters.

For the spin-singlet channel, the intra-band order parameter has the form
\begin{equation}
	\Delta^{\rm s}(\kk) = \psi(\kk) i \sigma_y \tau_0.
	\label{eq:singlet-gap}
\end{equation}
This order parameter describes fully-gapped s-wave spin-singlet pairing with $\psi(-\kk) = \psi(\kk)\equiv\psi$.

For the spin-triplet gap, Cooper pairing can be both within the plane and between neighboring planes: The first,
%$\Delta^{\rm t}(\kk) = \Delta^\parallel(\kk) + \Delta^\perp(\kk)$. For the in-plane pairing, the gap function has the general form
\begin{equation}
	\Delta^{\parallel}(\kk) = (\vec{d}_\kk^{\phantom{.}\parallel} \cdot \vec{\sigma}) i\sigma_y\tau_0 ,
	\label{eq:triplet-gap-para}
\end{equation}
where $\vec{d}_{\kk}^{\phantom{.}\parallel}=(d_{\kk}^x, d_{\kk}^y, 0)^{\rm T} \parallel \! \vec{f}_{\kk}$, is similar to the case of non-centrosymmetric superconductors~\cite{frigeri:2004, smidman:2017} and transforms with $A_{2u}$. This gap vanishes for $k_x=k_y=0$, such that for closed Fermi surfaces as found in the q3DI and 3D regime, the gap has point nodes. Notice that in the non-centrosymmetric case with point group $C_{4v}$, the gap functions of $A_{1g}$ and $A_{2u}$ symmetry mix, resulting instead in line nodes for closed Fermi surfaces and dominant triplet contribution. 

In addition the $A_{2u}$ order parameter with out-of-plane spin-triplet pairing is allowed due to the staggered nature of the mirror-symmetry breaking with a nearest-neighbor gap function
\begin{equation}
	\Delta^{\perp}(\kk) = [(\vec{d}^\perp_\kk  \sigma_z) i\sigma_y]\cdot \vec{\tau} .
	\label{eq:triplet-gap-perp}
\end{equation}
While we can in general write a gap function of this form that is entirely intra-band, we will for simplicity consider in the following $\delta t_z =0$, such that $\Delta^{\perp}(\kk) = (d_{\kk}^z \sigma_z)i\sigma_y \tau_1$ with $d_{\kk}^z \propto \sin k_z /2$ as the lowest-order basis function~\footnote{To lowest order, this would be $(d_\kk^\perp)_1 = \Delta^\perp \sin k_z \cos k_z/2 $ and $(d_\kk^\perp)_2 = \frac{t'}{t}\Delta^\perp \sin k_z \sin k_z/2$.}. This corresponds to an inter-layer pairing between nearest layers. This gap function has a line node for $k_z=0$, which can be removed by mixing in the other spin-triplet component of $A_{1u}$ symmetry, namely $\vec{d}_{\kk} = (\sin k_x \sigma_x + \sin k_y \sigma_y)i\sigma_y \tau_0$. This order parameter thus in general allows for a full gap.

While knowledge of the pairing interaction is needed to determine the leading instability in the system, we would expect the dominant channel to be in-plane in the quasi-two-dimensional limit. Consequently, the most probable order parameters are the spin-singlet and the spin-triplet of $A_{2u}$ symmetry. Only when approaching the three-dimensional limit, in other words $t_z \approx t$, an inter-layer pairing becomes feasible. As mentioned, this last pairing channel can lead to a full gap even in the three-dimensional limit, which in turn allows for topological superconductivity as we discuss in Sec.~\ref{sec:topology}.

%{\color{blue} Here a little discussion on when do we expect which one: In the quasi-two-dimensional limit, we certainly expect the intra-layer pairing to be the dominant. However, when we approach the 3D limit, the inter-layer pairing could become possible...}

%{\color{blue} we need to discuss quickly the nodal structure of the gaps here: The A1u I think can be fully gapped out. The A2u has probably either point nodes -- the simples case -- or even Bogoliubov Fermi surfaces... }
%Importantly, due to the out-of-plane contribution, an odd superconducting gap in the staggered-Rashba case does not have any nodes in the gap even for three-dimensional Fermi surfaces.
%Note that the full gap is an important difference compared to the 3D Rashba system without global inversion, where the spin-orbit coupling forces the $d$ vector to be parallel to $\vec{f}_{\kk}$ and a gap function with dominant spin-triplet contribution has in general line nodes on a three dimensional Fermi surface~\cite{brydon:2011c}.
 
\subsection{General spin response}

The spin response of a superconductor can be calculated similarly to the one in the normal state, Eq.~\eqref{eq:chi0}, using~\cite{abrikosov:1962}
\begin{multline}
  \chi_{ij} = -\mu_B^2 T \sum_n\sum_{\kk}\tr[\sigma_i\tau_0G(\kk, \omega_n)\sigma_{j}\tau_0G(\kk, \omega_n) \\
  - \sigma_i\tau_0F(\kk,\omega_n)\sigma_j^T\tau_0F^\dag(\kk, \omega_n)],
  \label{eq:chi}
\end{multline}
where $G(\kk, \omega_n)$ and $F(\kk, \omega_n)$ are the normal and anomalous Green's functions, respectively. Given the intra-band order parameters of Eqs.~\eqref{eq:singlet-gap}, \eqref{eq:triplet-gap-para}, or \eqref{eq:triplet-gap-perp}, these can be calculated explicitly using the Gor'kov equations, see App.~\ref{app:gorkov}. 
The resulting structure of both Green's functions is very similar to the normal state Green's function of the previous section, Eq.~\eqref{eq:normalgreens0}. In particular, the normal Green's function reads
\begin{equation}
  G(\kk, \omega_n)=G_{+}\sigma_0\tau_0 + G_{-}[\hat{\xi}_{\kk}^-\sigma_0\tau_{1}+\hat{f}_{\kk}\cdot\vec{\sigma} \tau_{3}]
  \label{eq:normalgreens}
\end{equation}
with
\begin{equation}
  G_{\pm}\! = G_{\pm}(\kk, \omega_n) =\!-\frac12\Big(\frac{i\omega_n +\xi_{+}}{\omega_n^{2}\!+E_+^2}\pm \frac{i\omega_n +\xi_{-}}{\omega_n^{2}\! + E_-^2}\Big)
  \label{eq:normalgpm}
\end{equation}
and $E_\alpha = \sqrt{\xi_\alpha^2 \! + \! |\Delta|^2}$.
The normal Green's function does not depend on the specifics of the intra-band gap function, other than its (momentum-dependent) magnitude, in other words $|\Delta|^2 =|\psi|^2$ for the spin-singlet, $|\Delta|^2 \equiv |\Delta^{\parallel}_\kk|^2$ for the intra-layer spin-triplet, and $|\Delta|^2 \equiv |\Delta^\perp_\kk|^2$ for the inter-layer spin-triplet case.

For the anomalous Green's function, on the other hand, the gap structure enters explicitly,
\begin{equation}
  F(\kk, \omega_n) = \Big\{F_+\tau_0 +  F_{-}[\hat{\xi}_{\kk}^-\sigma_0\tau_{1}+ \hat{f}_{\kk}\cdot\vec{\sigma} \tau_{3}]\Big\}\Delta(\kk)
  \label{eq:anomalousgreens}
\end{equation}
with
\begin{equation}
  F_{\pm} = F_{\pm}(\kk, \omega_n) = \frac12\Big(\frac{1}{\omega_n^{2}+E_+^2}\pm \frac{1}{\omega_n^{2} + E_-^2}\Big).
  \label{eq:anomalousgpm}
\end{equation}

To proceed, we separate the susceptibility in Eq.~\eqref{eq:chi} into a `normal' and an `anomalous' part, $\chi_{ij} = \chi^{\rm n}_{ij} + \chi^{\rm a}_{ij}$. The trace of the normal part is the same for all three order parameters and, as in the discussion in the previous section, yields
\begin{equation}
  \frac{\chi^{\rm n}_{z}(\kk, \omega_n)}{4\mu_B^2 T} = -\{(G_{+})^2 + (G_{-})^2 [(\hat{\xi}_{\kk}^-)^2-|\hat{f}_{\kk}|^2]\}
    \label{eq:tottraceGsz}
\end{equation}
and similarly
\begin{equation}
    \frac{\chi^{\rm n}_{x}(\kk, \omega_n)}{4\mu_B^2 T}  =  -\{(G_{+})^2 + (G_{-})^2 [(\hat{\xi}_{\kk}^-)^2+(\hat{f}_\kk^x)^2-(\hat{f}_\kk^y)^2]\}.
    \label{eq:tottraceGsx}
\end{equation}
As the gap structure enters the anomalous Green's function, we discuss in the following first the susceptibility for the case of the spin-singlet and then the case of the spin-triplet order parameters.

\subsubsection{Spin-singlet order parameter}
Using the $s$-wave singlet gap function in the anomalous Green's functions, Eq.~\eqref{eq:anomalousgreens}, we find for the anomalous part of the susceptibility
\begin{equation}
  \frac{\chi^{\rm a}_{z}(\kk, \omega_n)}{4\mu_B^2 T} = -|\Delta|^2 \{(F_{+})^2 + (F_{-})^2 [(\hat{\xi}_{\kk}^-)^2-|\hat{f}_{\kk}|^2]\}
    \label{eq:tottraceFsz}
\end{equation}
and similarly
\begin{equation}
    \frac{\chi^{\rm a}_{x}(\kk, \omega_n)}{4\mu_B^2 T}  =  -|\Delta|^2\{(F_{+})^2 + (F_{-})^2 [(\hat{\xi}_{\kk}^-)^2+(\hat{f}_\kk^x)^2-(\hat{f}_\kk^y)^2]\}.
    \label{eq:tottraceFsx}
\end{equation}
We can again separate the susceptibility into two contributions, namely
\begin{align}
	\chi_{\rm P}^+(\kk, \omega_n) &= (G_+)^2 + (G_-)^2 + |\Delta|^2 [(F_+)^2 + (F_-)^2]   \nonumber\\
	&=\frac12 \sum_{\alpha=\pm}\frac{(i\omega_n + \xi_{\alpha})^2 + |\Delta|^2}{(\omega_n^2 + E_\alpha^2)^2}
\end{align}
and
\begin{align}
	\chi_{\rm vV}^+ (\kk, \omega_n)&= (G_+)^2 - (G_-)^2 + |\Delta|^2 [(F_+)^2 - (F_-)^2]   \nonumber\\
	&= \frac12\Big\{\frac{(i\omega_n - \xi_{+})(i\omega_n - \xi_{-})+|\Delta|^2}{(\omega_n^2 + E_+^2)(\omega_n^2 + E_-^2)}\Big\}.
\end{align}
The total susceptibility after Matsubara summation is given by
\begin{equation}
  \chi^{\rm s}_{z} = \sum_{\kk}(\hat{\xi}_{\kk}^-)^2\chi^{+}_{\rm P}(\kk) + |\hat{f}_{\kk}|^2\chi^{+}_{\rm vV}(\kk)
  \label{eq:chizsinglet}
\end{equation}
for fields along $z$ and, for fields along the $x$ or $y$ direction, the susceptibility has the form
\begin{equation}
  \chi^{\rm s}_{x} =  \sum_{\kk}[(\hat{\xi}_{\kk}^-)^2 + (\hat{f}_\kk^x)^2]\chi^{+}_{\rm P}(\kk) + (\hat{f}_\kk^y)^2\chi^{+}_{\rm vV}(\kk).
  \label{eq:chixsinglet}
\end{equation}
The Matsubara sums are evaluated in App.~\ref{app:matsubara} and yield
\begin{equation}
  \chi^{+}_{\rm P}(\kk) = 2 \mu_B^2 \sum_{\alpha}\frac{1}{4T\cosh^2(E_{\alpha}/2T)}
  \label{eq:paulip}
\end{equation}
for the Pauli term and, for $\Delta \ll \sqrt{|\vec{\xi}_{\kk}^-|^2 + |\vec{f}_\kk|^2}$, we can approximate
\begin{equation}
  \chi^+_{\rm vV}(\kk) \approx 2\mu_B^2 \frac{[n_{\rm F}(E_{+}) - n_{\rm F}(E_{-})]}{\sqrta}\approx \chi_{\rm vV}^0(\kk)
  \label{eq:vanVleckP}
\end{equation}
for the van Vleck susceptibility.

\subsubsection{Intra-layer spin-triplet pairing}
For spin-triplet pairing, we start with the intra-layer-pairing case with order parameter $\Delta^{\parallel}(\kk) = (\vec{d}_{\kk}^{\phantom{.}\parallel} \cdot \vec{\sigma}) i\sigma_y\tau_0$, where $\vec{d}_{\kk}^{\phantom{.}\parallel} \parallel \vec{f}_{\kk}$. For out-of-plane fields, $i=j=z$, the anomalous part of the susceptibility yields
\begin{equation}
 \frac{\chi_{z}^{\rm a}}{4\mu_B^2 T} = |\Delta_{\kk}|^2 \{F_+^2 + F_-^2[(\hat{\xi}_{\kk}^-)^2-|\hat{f}_{\kk}|^2]\}.
  \label{eq:strfztriplet_intra}
\end{equation}
For in-plane fields, $i=j=x$, we find
\begin{equation}
  \frac{\chi_{x}^{\rm a}}{4 \mu_B^2 T} =  (|d_{\kk}^y|^2 - |d_{\kk}^x|^2)  \{F_+^2 + F_-^2 [(\hat{\xi}_{\kk}^-)^2+(\hat{f}_\kk^x)^2-(\hat{f}_\kk^y)^2]\}.
  \label{eq:strfxytriplet_intra}
\end{equation}
Combining Eqs.~\eqref{eq:tottraceGsz} and \eqref{eq:tottraceGsx} with \eqref{eq:strfztriplet_intra} and \eqref{eq:strfxytriplet_intra}, we thus find for the total susceptibility two additional contributions compared to the spin-singlet case. First,
\begin{align}
	\chi_{\rm P}^-(\kk, \omega_n) &= (G_+)^2 + (G_-)^2 - |\Delta|^2 [(F_+)^2 + (F_-)^2]   \nonumber\\
	&=\frac12 \sum_{\alpha=\pm}\frac{(i\omega_n + \xi_{\alpha})^2 - |\Delta|^2}{(\omega_n^2 + E_\alpha^2)^2},
	\label{eq:PauliM}
\end{align}
which,  after performing the Matsubara sum (App.~\ref{app:matsubara}) yields
\begin{multline}
  \chi^{-}_{\rm P}(\kk) = 2 \mu_B^2 \sum_{\alpha}\frac{|\Delta|^2}{E_\alpha^3}\tanh\Big(\frac{E_\alpha}{2T}\Big) \\+\frac{\xi_\alpha^2}{E_\alpha^2}\frac{1}{4T\cosh^2(E_\alpha/2T)}.
  \label{eq:paulim}
\end{multline}
The second,
\begin{align}
	\chi_{\rm vV}^- (\kk, \omega_n)&= (G_+)^2 - (G_-)^2 - |\Delta|^2 [(F_+)^2 - (F_-)^2]   \nonumber\\
	&= \frac12\Big\{\frac{(i\omega_n - \xi_{+})(i\omega_n - \xi_{-})-|\Delta|^2}{(\omega_n^2 + E_+^2)(\omega_n^2 + E_-^2)}\Big\}
	\label{eq:vanVleckM}
\end{align}
yields for $\sqrt{|\vec{\xi}_{\kk}^-|^2 + |\vec{f}_\kk|^2} \gg \Delta$ the same van Vleck contribution as Eq.~\eqref{eq:vanVleckP}, $\chi^-_{\rm vV}(\kk) \approx \chi^+_{\rm vV}(\kk)\equiv\chi_{\rm vV}(\kk)$.
For fields out of plane, we thus find the susceptibility
\begin{equation}
  \chi^{\rm t, \parallel}_{z} = \sum_{\kk}(\hat{\xi}_{\kk}^-)^2\chi^{-}_{\rm P}(\kk) + |\hat{f}_{\kk}|^2\chi_{\rm vV}(\kk)
  \label{eq:chiztriplet_intra}
\end{equation}
and for the susceptibility for in-plane fields
\begin{widetext}
\begin{equation}
  \chi^{\rm t, \parallel}_{x} = \sum_{\kk}\Big\{[(\hat{\xi}_{\kk}^-)^2 + (\hat{f}_\kk^x)^2]\frac{|d^x_{\kk}|^2}{|\Delta_{\kk}|^2}\chi^{+}_{\rm P}(\kk)+  \frac{|d_{\kk}^y|^2}{|\Delta_\kk|^2} \chi^{-}_{\rm P}(\kk)\Big]
  + (\hat{f}_\kk^y)^2\chi_{\rm vV}(\kk)\Big\}.
  \label{eq:chixtriplet_intra}
\end{equation}
\end{widetext}

\subsubsection{Inter-layer pairing}
For the inter-layer pairing with $\Delta^{\perp}(\kk) =(d_\kk^z \sigma_z) i \sigma_y \tau_1$, the traces of the anomalous part of the susceptibility in Eq.~\eqref{eq:chi} yield for out-of-plane fields
\begin{equation}
 \frac{\chi_{z}^{\rm a}}{4\mu_B^2 T} = - |d^z_{\kk}|^2 \{F_+^2 + F_-^2[(\hat{\xi}_{\kk}^-)^2-|\hat{f}_{\kk}|^2]\}
  \label{eq:strfztriplet_inter}
\end{equation}
and for in-plane fields, $i=j=x$, we find
\begin{equation}
  \frac{\chi_{x}^{\rm a}}{4 \mu_B^2 T} =  |d^z_{\kk}|^2 \{F_+^2 + F_-^2 [(\hat{\xi}_{\kk}^-)^2+(\hat{f}_\kk^x)^2-(\hat{f}_\kk^y)^2]\}.
  \label{eq:strfxytriplet_inter}
\end{equation}
For fields out of plane, we thus find the susceptibility
\begin{equation}
  \chi^{\rm t, \perp}_{z} = \sum_{\kk}(\hat{\xi}_{\kk}^-)^2\chi^{+}_{\rm P}(\kk) + |\hat{f}_{\kk}|^2\chi_{\rm vV}(\kk)
  \label{eq:chiztriplet_inter}
\end{equation}
and for in-plane fields the susceptibility 
\begin{equation}
  \chi^{\rm t, \perp}_{x} = \sum_{\kk}\Big\{[(\hat{\xi}_{\kk}^-)^2 + (\hat{f}_\kk^x)^2]\chi^{-}_{\rm P}(\kk)
  + (\hat{f}_\kk^y)^2\chi_{\rm vV}(\kk)\Big\}.
  \label{eq:chixtriplet_inter}
\end{equation}

\subsection{Discussion}

\begin{figure}[tt]
  \centering
  \includegraphics{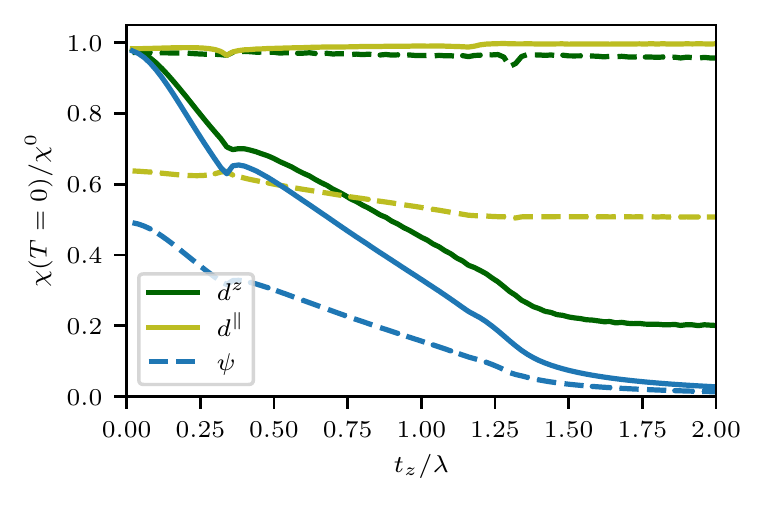}
  \caption{Out-of-plane (solid lines) and in-plane (dashed lines) susceptibility for all three order parameters for $T\rightarrow 0$ compared to the normal state susceptibility as a function of $z$ axis hopping for $\delta t_z =0$. The pairing gaps are defined as $\Delta(\kk) = \psi$ for the spin-singlet, $\vec{d}_{\kk}^\parallel = \Delta (-\sin k_y, \sin k_x, 0)$ for intra-layer pairing, and $d_{\kk}^z = \Delta \sin k_z / 2$ for inter-layer pairing. For the numerical evaluation, we used $\Delta = 0.025t$ and $T=0.0025t$.}
  \label{fig:chi_sc}
\end{figure}

Because the Pauli susceptibility in Eq.~\eqref{eq:paulip} vanishes for $T \to 0 $ for the spin-singlet state, only the temperature-independent van Vleck contribution of the normal state remains. Moreover, comparing Eqs.~\eqref{eq:chizsinglet} and \eqref{eq:chixsinglet}, we find for the spin-singlet pairing $\chi_z^{\rm s} = 2 \chi_x^{\rm s}$.
In order to discuss the behavior of the susceptibility for the triplet states in the $T\rightarrow 0$ limit, we rewrite the term in Eq.~\eqref{eq:paulim} as 
\begin{equation}
  \chi^{-}_{\rm P}(\kk) = 2 \mu_B^2 \sum_{\alpha}\frac{\partial}{\partial \xi}\Big[\frac{\xi_\alpha}{E_\alpha}\tanh\Big(\frac{E_\alpha}{2T}\Big)\Big] \approx \chi^0_{\rm P}(\kk)
  \label{eq:partialpaulim}
\end{equation}
which, for $T \ll \Delta \ll t$, reduces to a derivative of a step function around $\mu$ and, thus, approximately to the normal-state Pauli contribution to the susceptibility.
Hence, we find that the spin susceptibility for out-of-plane fields does not decrease for intra-layer pairing, while it follows the trend of the spin-singlet case for inter-layer pairing. For in-plane fields, however, the intra-layer pairing state reduces the spin susceptibility, while the inter-layer pairing state is not paramagnetically limited. These findings are, thus, in accordance with the general expectation that for a $d$-vector perpendicular to the magnetic field the susceptibility stays constant, while the contribution parallel to the field reduces the susceptibility. Note that here, the van Vleck term is generally not affected by superconductivity, since it arises form inter-band contributions, and the band splitting is governed by $\sqrt{|\vec{\xi}_{\kk}^-|^2 + |\vec{f}_\kk|^2} \gg \Delta$.

Figure~\ref{fig:chi_sc} shows the susceptibilities for all three order parameters as a function of the $z$-axis hopping $t_z$, where, as mentioned, we use $\delta t_z=0$ for simplicity. As discussed above, the zero-temperature in-plane susceptibility is always reduced to half of the out-of-plane susceptibility for (fully-gapped) spin-singlet pairing, while this is not the case for spin-triplet order parameters. Note that the inter-layer state has a larger susceptibility in these calculations due to the line nodes at $k_z = 0$. These nodes can in principle be gapped out, such that the out-of-plane susceptibility of the spin-singlet and the inter-layer spin-triplet would be equal. Finally, note that the in-plane (out-of-plane) pairing state is limited by orbital depairing for out-of-plane (in-plane) fields.

\section{Topological Considerations}
\label{sec:topology}
\begin{figure}[bb]
    \centering
    \includegraphics{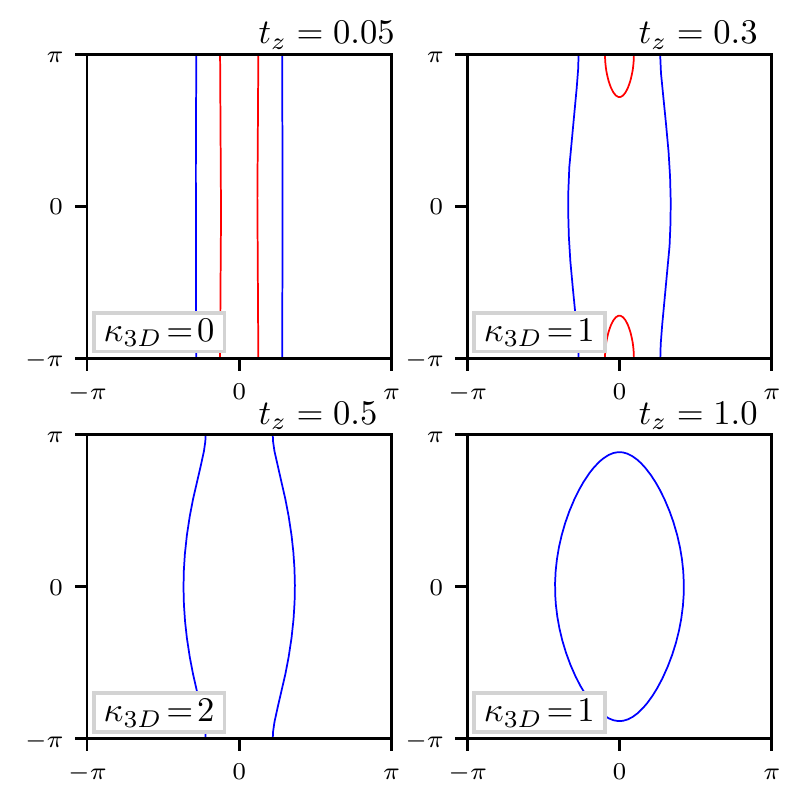}
    \caption{Projection of the Fermi surface onto $k_x$-$k_z$ plane for different values of interlayer hopping parameters $t_z$ and $\delta t_z$. Two colors correspond to two different momentum sectors with inversion eigenvalues respectively $+1$ and $-1$.}
    \label{fig:fermi_surface}
\end{figure} 

In this section, we investigate possible non-trivial topology of the superconducting phases in the different regimes identified in Sec.~\ref{sec:normal}.  
In systems invariant under inversion with an odd-parity superconducting order parameter, the topology of the superconducting phase can be identified using so-called symmetry indicators (SI)~\cite{skura:2020,ono:2020, geier:2020}. These indicators are defined in terms of the inversion eigenvalues of the occupied normal-state bands at time-reversal-symmetric momenta (TRIMs), where $-\kk \equiv \kk$ up to reciprocal lattice translations. In particular, the 3D inversion-symmetry indicator $\kappa_{3D}$ is a $\mathbb{Z}_8$-valued quantity, such that an odd SI indicates a strong topological superconductor, while $\kappa_{3D} = \pm2$ and $\kappa_{3D} = 4$  indicate second- and third-order topological phases, respectively.  

The inversion-symmetry indicator for a time-reversal symmetric odd-parity superconductor is defined as  
\begin{equation}
    \kappa_{3D} = \frac{1}{2}\sum_{{\bf{k}}\in \text{TRIMs}} \left( n^+_{{\bf k},\text{N}} - n^-_{{\bf k},\text{N}} \right ),
    \label{eq:indicator}
\end{equation}
where $n^{\pm}_{{\bf k},\text{N}}$ is the number of occupied bands at the TRIMs with inversion eigenvalue $\pm1$. 
To calculate the indicator, it is convenient to rewrite the dispersion in the $z$ direction, Eqs.~\eqref{eq:inter_layer1} and \eqref{eq:inter_layer2} as
\begin{align}
  (\tilde{\xi}_{\kk}^-)_1 &= - t_z + \delta t_z- (t_z + \delta t_z)\cos k_z\\
  (\tilde{\xi}_{\kk}^-)_2 &= - (t_z + \delta t_z)\sin k_z.
\end{align}
This corresponds to a gauge transformation such that the Fourier transform does not resolve the unit-cell structure anymore. The advantage, however, is that now $\mathcal{I} \HH_{\kk}  \mathcal{I}^{-1} = \HH_{-\kk} $ with $\mathcal{I}~=~\sigma_0\tau_1$ and we can straight-forwardly read off the inversion eigenvalues of the normal-state bands.

We are interested in inversion eigenvalues at two TRIMs $\Gamma = (0,0,0 ) $ and $Z = (0,0,\pi)$. At these two momenta, the two bands with energies given by Eq.~\eqref{eq:energies}, namely $\xi_{\Gamma,\pm} = -4t \mp 2 t_z$ and $\xi_{Z,\pm} = -4t \mp 2\delta t_z $, have inversion eigenvalues $\pm 1$. Depending on the Fermi-surface topology and hence, the number of bands occupied at the TRIMs, the system can realize a first- or second-order topological superconductor, see Fig.~\ref{fig:fermi_surface}. A third-order topological superconductor, however, is not possible at least with this simplified band structure, since for both bands occupied, the inversion eigenvalues are opposite and cancel in Eq.~\eqref{eq:indicator}. 

Table~\ref{tab:indicators} summarizes the symmetry indicators for the two spin-triplet (odd-parity) order parameters considered here. In particular, the $A_{2u}$ state, which pairs electrons within the layers and is, thus, more probable in the q2D limit, allows for a second-order phase when only one open Fermi surface exists. In the other potentially non-trivial cases, this gap structure has point nodes for $k_x=k_y=0$. The $A_{1u}$ state, which can be fully gapped for any Fermi-surface topology, thus allows in principle for both first- and second-order phases. As the $A_{1u}$ state corresponds to an anisotropic Balian-Werthammer state known from the B phase of ${}^3$He, a non-trivial topology might not be surprising~\cite{mizushima:2015}. Note, however, that this gap structure is rather unlikely in the quasi-2D limit.

\begin{table}[t]
    \centering
    \begin{tabular}{c|cccc}
        $\Delta$ & q2D & q3DI & q3DII & 3D\\
        \hline
        $A_{1u}$ & $(0)$ & $(1)$ & $2$ & $1$\\
        $A_{2u}$ & $0$ & $\times$ & $2$ & $\times$\\
    \end{tabular}
    \caption{Symmetry indicator for the two spin-triplet order parameters with $A_{1u}$ and $A_{2u}$ symmetry, where $1$ corresponds to a first-order and $2$ to a  second-order topological superconductor. The $\times$ indicates that the gap has point nodes and no strong topology is possible, while $(.)$ indicates that the inter-layer order parameter is not likely unless the system is three-dimensional.}
    \label{tab:indicators}
\end{table}

\section{Conclusion}
In some globally centrosymmetric systems, the lack of inversion symmetry in subunits can influence the physical properties significantly. Such remnants of non-centrosymmetricity are particular interesting in the context of superconductivity, where inversion symmetry yields a strict distinction between spin-singlet and spin-triplet Cooper pairing. In locally non-centrosymmetric systems, however, the two may mix in a very characteristic way.

In our work, we have investigated signatures of local noncentrosymmetricity for the case of a layered system with staggered layer-intrinsic Rashba spin-orbit coupling. Hereby, we have focused on different regimes going from a quasi-2D to fully 3D band structure upon tuning the interlayer hybridization. First looking at the normal state spin susceptibility, we identified four different regimes, characterized by their Fermi surface topology.

Unlike the case of globally non-centrosymmetric superconductors, where only one kind of spin-triplet state is feasible, we identified two such spin-triplet states in the staggered case. We analyzed the resulting three order parameters, namely a generic spin-singlet and those two spin-triplet phases and show that the comparison between their behavior under in-plane and out-of-plane magnetic fields allows for the order parameters' distinction. In particular, their low-temperature spin susceptibility displays different behavior that influence the paramagnetic limiting effects for different field orientations. 

For the spin-triplet order parameters, we have finally explored possible topological phases within the symmetry indicator framework, and identified both first- and second-order topological phases. Third-order topology is, on the other hand, not possible in the minimal model considered here.

Finally, we comment on the relevance of our results concerning CeRh$_2$As$_2$~\cite{khim:2021tmp}. For c-axis fields, the low-field state seems to be paramagnetically limited, which could have either spin-singlet or inter-layer spin-triplet character. However, as the in-plane fields show a smaller critical field for CeRh$_2$As$_2$, it is rather likely that the low-field phase constitutes spin-singlet pairing. Note that the spin-triplet phase that would in this scenario emerge at high fields, is certainly topologically trivial, as there are no first-order TSC in 3D. A generic possibility would, however, be a Weyl superconductor. As the field-induced phase transition indicates that the spin-triplet state is close by in parameter space, it might be possible to stabilize it also with other means, thus inducing possibly topological superconductivity.

%Again, the remaining trace is straight forward and yields
%\begin{equation}
%  \tr[\sigma_iF(\kk, \omega_n)\sigma_j^TF^\dag(\kk, \omega_n)] = - 4|d_{\kk}|^2 (F_{+}^2 + F_{-}^2)
%  \label{eq:ttrfztriplet}
%\end{equation}
%and
%\begin{equation}
%  \tr[\sigma_iF(\kk, \omega_n)\sigma_j^TF^\dag(\kk, \omega_n)] = 4|d_{\kk}|^2 (F_{+}^2 + F_{-}^2 - 2F_{-}^2 \frac{\ak^2}{|\ec|^2 + \ak^2})
%  \label{eq:ttrfxytriplet}
%\end{equation}
%\subsubsection*{The case $H || z$}
%This case look again exactly like the singlet case.. It should therefore also give the density of states\dots
%\subsubsection*{The case $H\perp z$}
%I still haven't really calculated this yet, I don't expect too big surprises, though\dots

\begin{acknowledgements}
We thank Daniel F Agterberg, Elena Hassinger, Seunghyun Khim, Titus Neupert, Eric Schertenleib, and Luka Trifunovic for fruitful discussions. A.S. was supported by funding from the European Research Council (ERC) under the European Union’s Horizon 2020 research and innovation program (ERC-StG-Neupert-757867-PARATOP) and by a Forschungskredit of the University of Zurich, grant No. FK-20-101. M.S. is financially supported by a Grant of the Swiss National Science Foundation (No.184739).
\end{acknowledgements}

\appendix

%%%%%%%%%%%%%%%%%%%%%%%%%%%%%%%%%%%%%%%%%%%%%%%%%%%%%%%%%%%%%%
\section{Gor'kov equations}\label{app:gorkov}
To calculate the Green's functions, we can use the Gor'kov equations
\begin{align}
  &G_0^{-1}(\kk, \omega_n) G(\kk, \omega_n) + \Delta_{\kk}F^\dag(\kk, \omega_n) = \tau_0\sigma_0\\
  (&G_0^{-1})^T(-\kk, -\omega_n) F^\dag(\kk, \omega_n) - \Delta^\dag_{\kk}G(\kk, \omega_n) = 0\label{eq:Gorkov2}\\
  &G_0^{-1}(\kk, \omega_n) F(\kk, \omega_n) - \Delta_{\kk}G^T(-\kk, -\omega_n) = 0.
  \label{eq:gorkov}
\end{align}
Equation~\eqref{eq:Gorkov2} leads to
\begin{equation}
  F^\dag(\kk, \omega_n) = G_0^T(-\kk, -\omega_n)\Delta^\dag_{\kk}G(\kk, \omega_n).
  \label{eq:fdag}
\end{equation}
We can use this in the first Gor'kov equation to find
\begin{equation}
  G^{-1}(\kk, \omega_n) = G_0^{-1}(\kk, \omega_n) + \Delta_{\kk}G^T_0(-\kk, -\omega_n) \Delta^\dag_{\kk}.
  \label{eq:g1}
\end{equation}
For the following, we thus need the normal state Green's function given by Eq.~\eqref{eq:normalgreens0}
and similarly
\begin{equation}
  G_0^T(-\kk, -\omega_n)=\tilde{G}^0_{+}\sigma_0\tau_0 + \tilde{G}^0_{-}[\hat{\xi}_{\kk}^-\sigma_0\tau_{3} + (\hat{f}_\kk^y \sigma_y - \hat{f}_\kk^x \sigma_x)\tau_{1}]
  \label{eq:gT}
\end{equation}
[$\tilde{G}^0_{\pm} = G^0_{\pm}(-\kk, -\omega_n)$ and $\vec{f}_{-\kk} = -\vec{f}_{\kk}$].
We are interested in the intra-band gap functions given in Eqs.~\eqref{eq:singlet-gap} or \eqref{eq:triplet-gap-para} and \eqref{eq:triplet-gap-perp}, for which we find
\begin{multline}
  \Delta_{\kk}G_0^T(-\kk, -\omega_n)\Delta^\dag_{\kk}  \\
   = |\Delta_\kk|^2 \Big[\tilde{G}^0_{+}\sigma_0\tau_0 + \tilde{G}^0_{-}[\hat{\xi}_{\kk}^-\sigma_0\tau_{3} + \hat{\vec{f}}_{\kk} \cdot \vec{\sigma}\tau_{1}]\Big]
  \label{eq:deltasinglet}
\end{multline}
with $\Delta^\pdag_{\kk} \Delta_{\kk}^\dag = |\Delta_{\kk}|^2 \sigma_0\tau_0$. Then
\begin{multline}
  G^{-1}(\kk, \omega_n) = (i\omega_n - \xi_{\kk}^+ + \Delta^2 \tilde{G}_+^0)\tau_0\sigma_0 \\
      +\Big(|\Delta_{\kk}|^2\tilde{G}^0_{-} - \sqrt{|\vec{f}_{\kk}|^2 + (\xi_{\kk}^-)^2}\Big)[\hat{\xi}_{\kk}^-\sigma_0\tau_{3} + \hat{\vec{f}}_{\kk} \cdot \vec{\sigma}\tau_{1}].
  \label{eq:Gm1}
\end{multline}
In order to invert this expression, we note that for each $\kk$, we can rotate the spins to be parallel to $z$ such that the inverse Green's function \eqref{eq:Gm1} has the structure
\begin{equation}
  G^{-1}(\kk, \omega_n) = A \tau_0 + B\hat{B}_s\cdot\tau
  \label{eq:Gm1b}
\end{equation}
with $\hat{B}_s$ a unit vector in the space of $\tau$ matrices for spin $s$. The inversion is then given by
\begin{equation}
	G(\kk, \omega_n) = G_+ \tau_0 + G_- \hat{B}_s\cdot \tau
\end{equation}
with
\begin{equation}
	G_{\pm}= \frac12 \left( \frac{1}{A+B} \pm \frac{1}{A-B}\right).
\end{equation}
Finally, we find
\begin{align}
  A\pm B &=  i\omega_n - \xi_{\kk}^+ + |\Delta_{\kk}|^2 (\tilde{G}_+^0\pm\tilde{G}_-^0) \mp \sqrt{|\vec{f}_{\kk}|^2 + (\xi_{\kk}^-)^2}\nonumber\\
  &= -\frac{\omega_n^2 + \xi_\pm^2 + |\Delta_{\kk}|^2}{i\omega_n + \xi_\pm},
  \label{eq:A1}
\end{align}
%and
%\begin{align}
%  A - B &=  i\omega_n - \xi_{kk}^0 + \Delta^2 (\tilde{G}_+^0-\tilde{G}_-^0) + \sqrt{|\vec{f}_{\kk}|^2 + (\xi_{\kk}^-)^2}\nonumber\\
%  &= -\frac{\omega_n^2 + \xi_-^2 + \Delta^2}{i\omega_n + \xi_-}
%  \label{eq:A1}
%\end{align}
such that, after rotating the spin back, the normal Green's function in the superconducting phase is given by Eq.~\eqref{eq:normalgreens}.
%\begin{equation}
%  G_{\pm} = -\frac12\Big(\frac{i\omega + \xi_+}{\omega_n^2 + \xi_+^2 + \Delta^2} \pm\frac{i\omega + \xi_-}{\omega_n^2 + \xi_-^2 + \Delta^2}\Big).
%  \label{eq:Greens}
%\end{equation}

Using Eq.~\eqref{eq:gorkov}, we can further calculate the anomalous Green's function to arrive at Eqs.~\eqref{eq:anomalousgreens} and \eqref{eq:anomalousgpm}.

\section{Matsubara sums of superconducting Green's functions}\label{app:matsubara}
The susceptibilities we calculated in the main text comprise four different terms of the general form denoted by $\chi^\pm_{\rm P}$ and $\chi^\pm_{\rm vV}$. In this appendix, we will explicitly perform the Matsubara sums to better understand these contributions.
%\begin{multline}
%	(G_+)^2 + (G_-)^2 \pm |\Delta_{\kk}|^2 [(F_+)^2 + (F_-)^2]   \\
%	=\frac12 \sum_{\lambda=\pm}\frac{(i\omega_n + \xi_{\lambda})^2 \pm |\Delta_{\kk}|^2}{(\omega_n^2 + \xi_{\lambda}^2 + |\Delta_{\kk}|^2)^2}
%	\label{eq:generalpauli}
%\end{multline}
%and
%\begin{multline}
%	(G_+)^2 - (G_-)^2 \pm |\Delta_{\kk}|^2 [(F_+)^2 - (F_-)^2]   \\= \frac12\Big\{\frac{(i\omega_n - \xi_{+})(i\omega_n - \xi_{-})\pm|\Delta_{\kk}|^2}{(\omega_n^2 + \xi_{+}^2 + |\Delta_{\kk}|^2)(\omega_n^2 + \xi_{-}^2 + |\Delta_{\kk}|^2)}\Big\}
%	\label{eq:generalvanvleck}
%\end{multline}
%We will refer to the former ones as Pauli terms and the latter ones as van Vleck terms, both of which we will discuss in the following.

\subsection{Pauli-like terms}
To evaluate the Matsubara sum of the first two contributions, Eqs.~\eqref{eq:paulip} and \eqref{eq:paulim}, we need to calculate the residues of
\begin{equation}
	 \frac{(z + \xi_{\lambda})^2 \pm |\Delta_{\kk}|^2}{(-z^2 + \xi_{\lambda}^2 + |\Delta_{\kk}|^2)^2}n_{\rm F}(z),
\end{equation}
at the singularities not stemming from the Fermi distribution function. We find two second-order poles, namely $z = \pm E = \pm \sqrt{\xi^2 + |\Delta|^2}$ (for simplicity, we omit the index $\alpha$ here). The residue is then
\begin{widetext}
\begin{align}
  \mathrm{Res}_{\pm E} &= \lim_{z\rightarrow \pm E} \frac{\partial}{\partial z} \Big[\frac{(z + \xi)^2 + |\Delta|^2}{(z \pm  E)^2}n_{\rm F}(z)\Big]\nonumber\\
	&= \lim_{z\rightarrow \pm E} \Big[\frac{2z + 2 \xi }{(z \pm  E)^2}n_{\rm F}(z) -  2\frac{(z + \xi)^2 + |\Delta|^2}{(z \pm  E)^3}n_{\rm F}(z) +\frac{(z + \xi)^2 + |\Delta|^2}{(z \pm  E)^2}n'_{\rm F}(z) \Big]\nonumber\\
	&= \Big[\frac{\pm 2E + 2 \xi }{4E^2} n_{\rm F}(\pm E)-  \frac{(\pm E + \xi)^2 + |\Delta|^2}{\pm 4E^3}n_{\rm F}(\pm E) +\frac{(\pm E + \xi)^2 + |\Delta|^2}{4E^2}n'_{\rm F}(\pm E) \Big] \nonumber\\
	&= \Big[\pm \frac{2E^2 \pm 2 E \xi  - E^2 \mp 2 E \xi - \xi^2 - |\Delta|^2}{4E^3}n_{\rm F}(\pm E) +\frac{(\pm E + \xi)^2 + |\Delta|^2}{4E^2}n'_{\rm F}(\pm E) \Big] \nonumber\\
	&= \Big[\pm \frac{E^2 - \xi^2 - |\Delta|^2}{4E^3}n_{\rm F}(\pm E) +\frac{(\pm E + \xi)^2 + |\Delta|^2}{4E^2}n'_{\rm F}(\pm E) \Big].
	\label{eq:residues}
\end{align}
\end{widetext}
Using that 
\begin{equation}
    n_{\rm F}'(z) = \frac{\partial n_{\rm F(z)}}{\partial z} = \frac{1}{4T\cosh^2(z/2T)}
    \label{eq:nfs}
\end{equation}
is an even function of $z$, we find for the sum of the two residues
\begin{equation}
	\mathrm{Res}_{E} + \mathrm{Res}_{-E} = \frac{1}{4T\cosh^2(E/2T)}.
\end{equation}
Similarly, replacing $|\Delta|^2$ with $-|\Delta|^2$ in Eq.~\eqref{eq:residues} for $\chi_{\rm P}^-$, we find
\begin{equation}
	\mathrm{Res}_{E} + \mathrm{Res}_{-E} = \frac{|\Delta|^2}{E^3}\tanh\Big(\frac{E}{2T}\Big) +\frac{\xi^2}{E^2}\frac{1}{4T\cosh^2(E/2T)}.
\end{equation}
Finally, we write for the two Pauli contributions
\begin{equation}
  \chi^{+}_{\rm P}(\kk) =  2 \mu_B^2 \sum_{\lambda}\frac{1}{4T\cosh^2(E_{\lambda}/2T)}
  \label{eq:pauliP}
\end{equation}
\begin{equation}
  \chi^{-}_{\rm P}(\kk) = 2 \mu_B^2 \sum_{\lambda}\frac{|\Delta|^2}{E_\lambda^3}\tanh\Big(\frac{E_\lambda}{2T}\Big) +\frac{\xi_\lambda^2}{E_\lambda^2}\frac{1}{4T\cosh^2(E_\lambda/2T)}.
  \label{eq:pauliMApp}
\end{equation}
%\begin{align}
%  \chi^{+}_{\rm P}(\kk) &=  -4\mu_B^2 T \sum_{n}\Big\{(G_{+})^2 + (G_{-})^2 + |\Delta|^2[(F_{+})^2 + (F_{-})^{2}]\Big\}\\
%  &= -2{\mu_B^2 T} \sum_{\alpha=\pm}\sum_{n}\frac{(i\omega_n + \xi_{\alpha})^2 + |\Delta|^2}{(\omega_n^2 + \xi_{\alpha}^2 + |\Delta|^2)^2}\\
%  &= 2 \mu_B^2 \sum_{\alpha}\frac{1}{4T\cosh^2(E_{\alpha}/2T)}.
%  \label{eq:pauliP}
%\end{align}
%\begin{align}
%  \chi^{-}_{\rm P}(\kk) &= -4\mu_B^2 T \sum_{n}\Big\{(G_{+})^2 + (G_{-})^2 - |\Delta|^2[(F_{+})^2 + (F_{-})^{2}]\Big\}\\
%  &= -2{\mu_B^2 T} \sum_{\alpha=\pm}\sum_{n}\frac{(i\omega_n + \xi_{\alpha})^2 - |\Delta|^2}{(\omega_n^2 + \xi_{\alpha}^2 + |\Delta|^2)^2}\\
%  &= 2 \mu_B^2 \sum_{\alpha}\frac{|\Delta|^2}{E_\alpha^3}\tanh\Big(\frac{E_\alpha}{2T}\Big) +\frac{\xi_\alpha^2}{E_\alpha^2}\frac{1}{4T\cosh^2(E_\alpha/2T)}
%  \label{eq:pauliM}
%\end{align}
The terms with the $\cosh$ vanish for $T\rightarrow 0$, such that only the first term of the latter equation survives,

\subsection{van Vleck--like terms}
The second set of contributions, Eqs.~\eqref{eq:vanVleckP} and \eqref{eq:vanVleckM}, has four poles at $\pm E_{\pm}$ and they are all first order. At $T=0$, the poles with positive energy thus lead to a vanishing Fermi function while the Fermi functions for the negative energies yield 1. We thus find
\begin{equation}
  \chi_{\rm vV}^{\pm}(\kk)=2\mu_B^2\sum_{\alpha=\pm}\alpha\frac{(E_{\alpha} + \xi_{+})(E_{\alpha} +\xi_{-}) \pm |\Delta|^2} {2E_{\alpha}(E_{+}-E_{-})(E_{+} + E_{-})}
  \label{eq:mats2}
\end{equation}
The energy scale for the van Vleck susceptibility is given by $\sqrta$ and since we assume $\sqrta\gg\Delta$, we proceed by setting the gap to zero. Then, we find
\begin{equation}
\begin{split}
  \chi_{\rm vV}^{\pm}(\kk)\approx \chi_{\rm vV}(\kk) &= 2\mu_B^2\Big[\frac{(|\xi_{+}| + \xi_{+})(|\xi_{+}| +\xi_{-})} {2|\xi_{+}|\xi^+_\kk\sqrta} \\
  &- \frac{(|\xi_{-}| + \xi_{+})(|\xi_{-}| +\xi_{-})} {2|\xi_{-}|\xi^+_\kk\sqrta}\Big]
  \label{eq:almost}
  \end{split}
\end{equation}
This expression yields exactly zero if both, $\xi_{+}$ and $\xi_{-}$, are either positive or negative. However, when only one of them is negative, while the other is positive, i.e. $n_F(\xi_+) - n_F(\xi_-)\neq0$, we find
\begin{equation}
  \chi_{\rm vV}(\kk) \approx 2\mu_B^2 \frac{[n_{F}(\xi_{+}) - n_F(\xi_{-})]}{\sqrta}
  \label{eq:zeroT}
\end{equation}
the van Vleck susceptibility.

\bibliography{refs}
\end{document}